\documentclass[useAMS,usenatbib]{mn2e}
\usepackage{amsmath,epsf}

\def\lesssim{\mathrel{\hbox{\rlap{\hbox{\lower4pt\hbox{$\sim$}}}\hbox{$<$}}}}
\def\gtrsim{\mathrel{\hbox{\rlap{\hbox{\lower4pt\hbox{$\sim$}}}\hbox{$>$}}}}
\newcommand{\ltaraw}{$\; \buildrel < \over \sim \;$}
\newcommand{\lta}{\lower.5ex\hbox{\ltaraw}}
\newcommand{\gtaraw}{$\; \buildrel > \over \sim \;$}
\newcommand{\gta}{\lower.5ex\hbox{\gtaraw}}
\def\lsim{\mathrel{\rlap{\lower4pt\hbox{\hskip1pt$\sim$}}
    \raise1pt\hbox{$<$}}}                % less than or approx. symbol

\newcommand{\ie}{{\it i.e.~}}
\newcommand{\eg}{{\it e.g.~}}

\newcommand{\etal}{{\it et~al.~}}
\newcommand{\cf}{{\it c.f.~}}

\newcommand{\Chandra}{{\it Chandra}}
\newcommand{\XMM}{{\bf XMM}}

\newcommand{\LT}{$L_x-T_x$}
\newcommand{\MT}{$M_t-T_x$}

\def\araa{ARA\&A}%
\def\apj{ApJ}%
\def\apjl{ApJ}%
\def\aap{A\&A}%
\def\mnras{MNRAS}%
\def\nat{Nature}%
\newcommand{\rc}{$r_c$}
\newcommand{\DT}{${\Delta}_T$}
\newcommand{\betarc}{$\beta-r_c$}
\newcommand{\DTrc}{${\Delta}_T-r_c$}

\newcommand{\trelax}{$t_{relax}$}

\newcommand{\taccrete}{$t_{accrete}$}
\newcommand{\tclosest}{$t_{closest}$}
\newcommand{\tapo}{$t_{apo}$}

\newcommand{\kpc}{\,\mbox{kpc}}

\newcommand{\Gyr}{\,\mbox{Gyr}}
\newcommand{\Myr}{\,\mbox{Myr}}

\newcommand{\keVcmsq}{\,\mbox{keVcm$^{2}$}}
\newcommand{\Sforty}{\,\mbox{$S_{40}$}}
\newcommand{\ks}{\,\mbox{ks}}

\newcommand{\Gyrs}{\,\mbox{Gyrs}}

\title[The impact of mergers on relaxed X-ray clusters III]
  {The impact of mergers on relaxed X-ray clusters \\
   III. Effects on compact cool cores}

\author[Gregory B. Poole \etal]{Gregory B. Poole$^{1}$\thanks{E-mail: gpoole@astro.swin.edu.au}, Arif Babul$^2$, Ian G. McCarthy$^3$, \newauthor A. J. R. Sanderson$^4$ and Mark A. Fardal$^5$ \\
  $^1$ Centre for Astrophysics and Supercomputing, Swinburne University of Technology, P.O. Box 218, Hawthorn, VIC 3122, Australia\\
  $^2$ Dept. of Physics \& Astronomy, University of Victoria, Elliott Building, 3800 Finnerty Rd., Victoria, BC, V8P 1A1, Canada \\
  $^3$ Department of Physics, University of Durham, South Road, Durham DH1 3LE, UK\\
  $^4$ School of Physics and Astronomy, University of Birmingham, Edgbaston, Birmingham, B15 2TT\\
  $^5$ Dept. of Astronomy, University of Massachusetts, Amherst, MA, 01003, USA}

\date{draft version \today}
\pagerange{\pageref{firstpage}--\pageref{lastpage}} \pubyear{2005}

\def\LaTeX{L\kern-.36em\raise.3ex\hbox{a}\kern-.15em
    T\kern-.1667em\lower.7ex\hbox{E}\kern-.125emX}

\begin{document}

\label{firstpage}

\maketitle

%-- ABSTRACT --
\begin{abstract}
We use the simulations presented in \citet{P06} to examine the effects of mergers on the properties of cool cores in X-ray clusters.  Motivated by recent \Chandra\ and \XMM\ observations, we propose a scheme for classifying the morphology of clusters based on their surface brightness and entropy profiles.  Three dominant morphologies emerge: two hosting compact cores and central temperatures which are cool (CCC systems) or warm (CWC systems) and one hosting extended cores which are warm (EWC systems).  In the cases we have studied, CCC states are disrupted only after \emph{direct} collisions with merging cluster cores.  This can happen in head-on collisions or during second pericentric passage in off-axis mergers.  By the time they are relaxed, our remnant cores have generally been heated to warm core (CWC or EWC) states but subsequently recover CCC states by the end of the simulation.  The only case resulting in a long-lived EWC state is a slightly off-axis 3:1 merger for which the majority of shock heating occurs during the accretion of a low-entropy stream formed from the disruption of the secondary's cool core.  Since $t_{dyn}\ll t_{cool}$ for all our relaxing merger remnant cores, compression prevents their core temperatures from falling until after they relax to the compact states allowed by their remnant central entropies.  This naturally explains the population of relaxed CWC systems observed in recent \Chandra\ and \XMM\ observations with no need to invoke AGN feedback.  The morphological segregation in the $L_x-T_x$ and $\beta-r_c$ scaling relations noted by \citet{MBBPH} are qualitatively reflected in the results of our mergers as well.  However, none of the cases we have studied produce systems with sufficiently high central entropies to account for the most under-luminous EWC systems observed.  Lastly, mergers do not efficiently mix the ICM.  As a result, merging systems which initially host central metallicity gradients do not yield merger remnants with flat metallicity profiles.  Taken together, these results suggest that once formed, compact core systems are remarkably stable against disruption from mergers.  It remains to be demonstrated exactly how the sizable observed population of extended core systems was formed.

\end{abstract}

%-- KEY WORDS --

%-- INTRO --
\section{Introduction}\label{sec-intro} 
The latest generation of X-ray telescopes have revolutionized our understanding of the intracluster medium (ICM) of galaxy clusters but yet one of the oldest observations of the field remains poorly understood.  As significant sample sizes of X-ray clusters initially became available, it was quickly realized that, broadly speaking, two populations could be identified from the shape of a system's X-ray surface brightness profile \citep{JonesForman84,OtaMitsuda04}: one with highly peaked central surface brightnesses (compact cores; $r_c\sim 50$\kpc) and one with flat central surface brightnesses (extended cores; $r_c\sim 150$ \kpc).  It was quickly understood that high and centrally peaked gas densities are primarily responsible for the central surface brightnesses of compact cores while extended cores possess lower and more constant central gas densities.  

As a consequence of the $\rho^2$ dependence of the ICM's emissivity, radiative cooling should be significantly more relevant to the structure of compact cores.  This assertion is supported by observations which have indicated that compact cores tend to host central positive temperature gradients \citep{DeGrandiandMolendi02}, low central entropies \citep{MBBPH} and radiative cooling rates which should release their thermal energy on timescales (the ``cooling time''; $t_{cool}$ hereafter) much shorter than the mean age of clusters \citep{Allen98}.  Within the context of specific physical models, these systems were once thought to host bulk radial flows of cold gas arising from the loss of pressure support their short cooling timescales induce.  As a result, they were classically referred to as ``cooling-flow'' systems.  Extended cores on the other hand tend to have flat central temperature profiles, elevated central entropies and cooling times in excess of a Hubble time.  Classically, these systems were referred to as ``non-cooling flow'' systems.

Great observational progress has been made since the launch of \Chandra~ and \XMM\ and cluster cores are now understood to be much more complicated than previously believed.  For example, it has been shown that clusters with compact cores can in fact have a range of central temperatures -- some significantly cooler than their surroundings and some roughly isothermal \citep{Donahueetal05,Sandersonetal06,Prattetal06}.  Hence, compact cores are not necessarily cool.  It is presently unclear what relative roles active galactic nuclei (AGN), mergers and cooling play in setting this spectrum of states.

As a result, the terminology of the field is in flux.  Given this situation, we propose here the adoption of new nomenclature for describing the morphology of X-ray cluster cores.   Anticipating the results to be presented in this paper, we will refer to extended core clusters as extended warm core (EWC) systems and compact core clusters as either compact cool core (CCC) or compact warm core (CWC) systems.  We shall illustrate in this paper why this facilitates a more insightful connection to the physical processes shaping the structure of cluster cores.

Despite recent progress in understanding the structure of cluster cores, the origin and precise nature of their observed distribution of surface brightness core radii still lacks a clear and definitive explanation.  Several hypotheses have been proposed to account for these populations, with three being particularly persuasive.

First, it has been known for a decade that radiative cooling and shock heating during accretion are not sufficient to account for the observed structure of X-ray clusters.  \citet[][M04 henceforth]{MBBPH} have developed a model in which entropy injection (possibly due to galactic outflows or early AGN activity) during the protocluster phase establishes a spectrum of minimum entropies for clusters leading to variations in cluster core cooling efficiencies.  Reasonable amounts of such additional entropy ($\sim200$ \keVcmsq) and radiative cooling can account for the normalization and scatter in the observed \LT~ and \MT~ relations while obeying observed cold gas constraints.  M04 also illustrate how clusters receiving little entropy injection quickly evolve to states resembling compact cool core systems as a natural consequence of radiative cooling.  Systems receiving higher levels remain in long lived extended core states.  Although not stated explicitly in their discussion, systems evolving to compact cool core states in this model would then be stabilized through several possible feedback processes involving late AGN activity \citep{VoitDonahue05,Nusseretal06,McCarthyetal07}.

Second, the entropy injection required to account for the structure of extended cores may have occurred more recently through interactions of AGN with the ICM.  Such interactions have been observed in the cores of several systems where jets from AGN are seen to be actively inflating large high-entropy cavities (or ``bubbles'').  It has been shown that the power supplied by such jets correlates with the local Bondi accretion rate and is energetically capable of offsetting the present-day effects of cooling \citep{Allenetal06,Bensonetal03,Crotonetal05,Boweretal06,Nemmenetal07}.  Typically, such interactions are localized to the central $\sim50$\kpc~ of a system but in a few instances, much larger outbursts reaching to $\sim200$\kpc~ have been observed \citep{McNameraetal05}.  However, as pointed out by \citet{McCarthyetal07}, present-day AGN activity is energetically capable of merely maintaining the present day configuration of actively cooling cluster cores.  The energy required to transform such systems into the observed range of non-cooling cores is much greater than the largest AGN outbursts yet observed.  This fact strongly suggests that an early epoch of "pre-heating" due to a yet unspecified mechanism (possibly AGN) is largely responsible for setting the observed range of present-day cluster core states.  Unfortunately, the short timescales, complicated magnetohydrodynamic physics and large range of relevant physical scales involved in the interaction of jets with the ICM makes exploring such issues from theoretical directions extremely challenging \citep{Churazovetal01,Ruszkowskietal04,DallaVecchiaetal04,Ommaetal04,Bruggenetal05,Vernaleoetal06,Heinzetal06,CattaneoTeyssier07}.

Lastly, mergers between clusters offer a compelling means by which to transform compact core systems into extended core systems.  It has been a tacit belief for some time that extended core systems are a product of merger interactions but the issue has received remarkably little detailed consideration.  This hypothesis is primarily based on early anecdotal observations that few compact cores exhibited disturbed morphologies while the limited number of studied extended core systems tended to be disturbed \citep{Edge92,BuoteTsaiII}.  However, with improved instrumentation and larger sample sizes, there are now several examples of disturbed compact core systems \citep[\eg A2204,][]{Sandersetal05}, including ones presently undergoing significant mergers \citep[\eg A2142,][]{Markevitchetal00}, as well as extended core systems lacking any evidence of having been disturbed at all (\eg\ Abell 2034 and Abell 2631; K. Cavagnolo, private communication).

Despite such concerns, the merger-origin hypothesis for extended core systems is appealing on theoretical grounds.  The presently favored hierarchical clustering paradigm for structure formation predicts that large objects are formed through regular accretion of smaller subclusters with each observed system representing the product of a unique history of many such events.  This process is seen to be active to the present epoch with recent \Chandra~ and \XMM~ observations routinely detecting evidence of significant recent or ongoing mergers.  With as much as $10^{64}$ ergs of initial kinetic energy, mergers are energetically capable of significantly modifying the structure of a system's ICM through the shocks they produce.  Furthermore, the wide variety of individual accretion histories may lead naturally to a range of remnant states like that described by M04.

Recent studies that have investigated cluster mergers in a context capable of testing the merit of this scenario have yielded inconclusive results.  In the influential study of \citet[][RS01 hereafter]{RickerandSarazin01} for instance, some evidence for the flattening of central density profiles due to shock heating from mergers is presented.  However, since the focus of their study was on merger induced luminosity and temperature ``boosts'', they did not include the effects of cooling (which are known to be important to the structure of compact core systems) and they present relatively little detailed analysis of their remnant cores' structures.  \citet{Gomezetal} and \citet{RT02} have constructed suites of idealized merging systems but their conclusions regarding the stability of compact core systems to mergers seem to conflict.  \citet{Gomezetal} find that systems with initially short cooling times will likely reestablish their structure while \citet{RT02} find that compact core systems are completely disrupted in equal mass and near-axis 8:1 mergers.  Adding to the controversy, studies of clusters produced in cosmological hydrodynamic simulations routinely fail to yield enough scatter in their X-ray scaling relations to account for observations \citep[\eg ][]{Rowleyetal04,Kayetal07}.

To address this and other gaps in our understanding of cluster mergers, we have initiated a numerical study of their impact on the observable properties of relaxed cool core X-ray clusters using idealized two-body simulations.  In a previous paper \citep[Poole06 hereafter]{P06} we have presented the details of our numerical approach and the dynamical evolution of a suite of 9 simulations (three mass ratios with three impact parameters each).  We then used these simulations for a second study examining the effects of mergers on global X-ray and SZ observables and their scaling relations \citep[][Poole07 hereafter]{P07}.  The same suite of simulations form the basis of the study presented in this paper.  We briefly review several aspects of our simulations particularly relevant to this paper in Section \ref{sec-method} and refer the reader to Poole06 for more details.

In this study we establish criterion for quantifying the evolving morphology of merging systems (Section \ref{analysis-morphology}) and present their evolution through three distinct morphological states (Section \ref{analysis-states}).  We then relate the occurrence of these states to changes in the structure of the system's core (Section \ref{analysis-structure}).  After considering the morphological segregation mergers induce in observed scaling relations (Section \ref{analysis-segregation}) we then consider the mixing effects of mergers on metallicity gradients (Section \ref{analysis-mixing}).  We then discuss several consequences of our findings in Section \ref{sec-discussion} and finally, summarize our study in Section \ref{sec-summary}.

In all cases our assumed cosmology will be ($\Omega_M$,$\Omega_\Lambda$)=($0.3$,$0.7$) with $H_o$=$75 km s^{-1} Mpc^{-1}$.

%-- INITIAL CONDITIONS AND NUMERICS --
\section{Simulations}\label{sec-method}

We run our simulations with GASOLINE \citep{wadsley04}, a versatile parallel SPH tree-code with multi-stepping.  We include the effects of radiative cooling in our simulations and feedback from star formation but the effects of jets from active galactic nuclei (AGN) are omitted.  

The basis of our study is a set of 9 idealized 2-body cluster merger simulations initialized to accurately match the properties of observed compact cool core clusters.  Three mass ratios are examined (1:1, 3:1 and 10:1) with the most massive system set to have a mass of $10^{15}M_{\odot}$ in every case.  For each mass ratio, three impact parameters are examined (one head-on and two off-axis cases).  We parametrize impact parameters by the smaller system's transverse velocity when its core crosses the virial radius of the massive system, in units of the circular velocity of the massive system at that radius ($v_t/V_c$).

In Poole06 we present a detailed account of how our isolated clusters and their orbits are initialized as well as the numerical methods and code parameters of our simulations.  In this section we summarize key aspects of our approach pertinent to our present analysis and direct readers looking for additional details to Poole06.

\subsection{Initial conditions}\label{numerics-cluster_initial}

To initialize the structure of our systems, we have followed the analytic prescription of \citet{BBLP} and \citet{MBBPH} to produce cool core clusters which conform with recent theoretical and observational insights into cluster structure.  The dark matter density profiles of our systems follow an NFW-like form \citep{NFW96,Mooreetal98} with the central asymptotic logarithmic slope chosen to be $\beta=1.4$ and the scale radius ($r_s$) selected to yield a concentration $c=R_{200}/r_s=2.6$ (we will use $R_{\Delta}$ throughout to indicate the radius within which the mean density of the system is $\Delta$ times the critical density, $\rho_c=3 H_0^2/8 \pi G$; $R_{200}=1785$\kpc, $R_{500}=1166$\kpc\ and $R_{cool}=180$\kpc\ initially for our most massive systems).  The initial gas density and temperature profiles of the clusters are set by requiring that (1) the gas be in hydrostatic equilibrium within the halo, (2) the ratio of gas mass to dark matter mass within the virial radius be $\Omega_b/(\Omega_m - \Omega_b)$, and (3) the initial gas entropy$^6$\footnotetext[6]{We use the standard proxy for entropy given by $S\equiv kT/n_e^{2/3}$ with $n_e$ and $T$ representing the electron density and temperature of the gas.} scale as  $S(r) \propto r^{1.1}$ over the bulk of the cluster body, as found in cosmological hydrodynamic simulations and observations of the ICM \citep{Lewisetal00,Voitetal05,Donahueetal06}.  We normalize the entropy profiles such that the temperature of the ICM at $R_{vir}$ is half the virial temperature.

We construct orbits for our systems which produce specified radial and tangential velocities for the secondary system ($v_r$ and $v_t$ respectively) when its centre of mass reaches the virial radius of the primary ($R_{vir}$).  For each of the three mass ratios we study, we examine three orbits selected to produce a typical value of $v_r(R_{vir})$ and to cover a significant range of the transverse velocity $v_t(R_{vir})$ giving rise to mergers found in cosmological dark matter simulations \citep{tormen97,vitvitska02}.  

Throughout our analysis we shall distinguish the two merging systems by calling the more massive system the ``primary'' system (we arbitrarily choose one to be the primary in the 1:1 cases, with no consequences for our results given their high degree of symmetry) and the less massive system as the ``secondary''.

\subsection{Dynamical evolution}\label{sec-dynamics}

In Poole06 we illustrated the generic dynamical progression which each of our simulations proceed through, identifying five distinct stages: a pre-interaction phase, first core-core interaction, apocentric passage, secondary core accretion, and relaxation.  These stages feature prominently in the discussion which follows so we will briefly review their progression here, for those who have not read Poole06.

As the cores accelerate towards each other during pre-interaction, a pair of shock fronts materialize and are driven towards each core, heating and compressing them briefly.  The core of the secondary system crosses within $R_{200}$ of the primary system at $t_o$ and at \tclosest\ they reach closest approach (\ie\ first pericentric passage).  In every case (including the head-on collisions), some part of the secondary's core survives its first encounter with the primary, tidally stripped into large cool streamers in off-axis cases and strings of several clumps in the on-axis cases.  At \tapo~ the disturbed cores reach maximum separation.  After subsequently reaching second pericentric passage at \taccrete, no observable trace of the secondary core remains.  This period marks the beginning of a prolonged period of accretion when the plume of material generated from the disruption of the secondary's core accretes as a high velocity stream.  This instigates a second episode of core heating followed by a period of relaxation.  At approximately \trelax , the system exhibits no obvious substructure in simulated $50$ks \Chandra~ images with the system placed at $z=0.1$ (see Poole06 for sample images and details on how this is done).

\subsection{Consequences of our coarsely sampled initial conditions}\label{numerics-statistics}

In this paper, we seek to understand the role which mergers are
playing in shaping the observed diversity of cluster core
morphologies.  It is important to note however that our simulations
are certainly not uniformly represented in the statistics of the
population.  In Poole07 we used merger rate statistics from
cosmological N-body simulations to argue that our 1:1 mergers are
expected to be rare, our 3:1 mergers uncommon but of statistical
significance and our 10:1 mergers extremely common (see Poole07 for a
more quantitative account).  For these reasons, when considering the
effects of mergers in the context of statistical properties of cluster
populations, we will be primarily concerned with the effects of 3:1 and 10:1 events.

We will find in later sections that the structure of cluster cores following mergers is the result of a delicate balance between shock heating and radiative cooling.  In particular, we shall illustrate in Section \ref{analysis-entropy} the important role stream accretion (see Poole06) plays in setting the entropy of our merger remnant's cores.  The efficiency of this process depends sensitively on many factors.  These include the mass and density of material present in the stream and the velocity it carries to the remnant core.  Both of these depend on the initial mass and orbit of the secondary in a complicated way, through (for example) their influence on the secondary core's disruption and the amount of dynamical friction experienced by the system.    

The impact parameters and mass ratios of our simulations were selected to represent the most interesting range of situations exhibited by dark matter substructure in cosmological simulations.  However, these ranges are sampled very sparsely by our simulations.   As a result, we do not claim to have performed a systematic analysis of many issues discussed in this paper.  We shall merely seek to reveal the general processes most relevant to establishing the structure of cluster cores following mergers.  Many details will have to be left open to future study.

\subsection{A cautionary note}\label{numerics-cores}

In SPH cluster simulations, cooling converts hot gas in the centre of the system into dense cold gas which is then (if star formation is included) transformed into a collisionless stellar medium.  As a result of the finite timescales involved in transforming cold gas into stars, a multiphase interface forms between dense cold gas and hot rarefied gas at the center.  It is well known that multiphase interfaces of this sort are poorly treated in SPH simulations \citep{RT01,MarriWhite03}.  Since the dynamical time in the center is much shorter than the cooling time, the result is a quasi-adiabatic flow which drives compressional heating as hot material collects several kernel radii outside of the multiphase interface.  This leads to artificially suppressed densities and enhanced temperatures and entropies in the cores.  The pooling of baryons to the center also leads to deeper central potentials which contribute to higher core temperatures.  During steady undisturbed cooling in our simulations, these effects appear initially at $r\sim 3$\kpc~ but slowly propagate outwards with time.

Such numerical concerns are significant only while a core is dense and its cooling timescales very short.  In each of our merger simulations, the cores of both systems become sufficiently disturbed by the interaction to restrict periods of active quiescent cooling to durations shorter than $4$\Gyrs.  To determine the physical extent of the regions influenced by this artifact, we have run our systems in isolation at both our nominal resolution and with $4\times$ its mass resolution.  After $4$\Gyrs, we find that the density and temperature profiles of these two simulations agree to within $10$\% outside of $30$\kpc.  Thus, the influence of such numerical effects should remain confined to radii $<30$\kpc,  marginalizing these concerns for our study.

To ensure that concerns which might arise as a result of these effects
are minimized, we will exclude the central $30$\kpc\ when
studying projected quantities (such as surface brightnesses and
central temperatures) and measure ``central'' quantities (such as
entropy) outside this radius, at $40$\kpc.  At a redshift of $z=0.1$,
$30$\kpc\ corresponds to $17$''; comparable to the typical central bin sizes
of \Chandra\ temperature profiles or the ROSAT
PSF ($\sim 15$'').  Since these are the instruments used to obtain the
data against which we compare our results, removal of this region should not introduce significant
biases during our comparisons to observations.

Since the effect of rising central temperatures is to reduce central temperature gradients, timescales of the recovery of warm core to cool core morphologies are likely exaggerated in our simulations, leading to systematic biases in our analysis.  It is critical to note however that the direction of this bias is always the same: towards overestimating the longevity of warm core morphologies.  Hence, the broad understanding and conclusions which emerge from this study are preserved.

%-- ANALYSIS --
\section{Analysis}\label{sec-analysis}
In this section we present our methods of classifying the evolving morphologies of our simulated mergers.  We shall find that as our mergers progress, they exist primarily in one of three states: compact cool core (CCC), compact warm core (CWC) and extended warm core (EWC) morphologies.  In the sections which then follow, we shall discuss the temporal evolution of our mergers through these three states and relate their occurrence to the changing structure of our merger remnants' cores.

\subsection{Observational Datasets}

To compare our classified systems to observations, we will employ the 20
cluster statistical sample presented by \citet{Sandersonetal06}.  We
have also augmented the results of this study by fitting 1 dimensional
$\beta$-models (\cf\ Equation 1 below) to each cluster.  To do so, we took the
APEC spectral normalization per unit area as a measure of projected
emissivity, using the more finely binned, projected annular spectral
profiles of \citet{Sandersonetal06} to construct azimuthally-averaged radial
surface brightness profiles. Eqn. 1 of \citet{Sandersonetal06}  was fitted using the
$\chi^2$ statistic to each profile across the full radial range to
determine the core radius; no attempt was made to excise or model any
excess emission associated with a cool core. The use of projected
emissivity instead of actual X-ray surface brightness is advantageous,
since it allows for the effects of galactic absorption, which is a
free-fitted parameter in each annular spectrum for most of the
clusters analyzed \citep[but fixed at the level expected from HI
surveys in the few clusters where this parameter was not well
constrained; see][for details]{Sandersonetal06}. For example, Abell 478,
suffers from strong galactic absorption with significant spatial
variation \citep{Sandersonetal05} that could bias fits based on the surface brightness.

In later sections we also employ the catalogue of ROSAT and ASCA observations presented by \citet{Horner_thesis}.  This catalogue is best suited to comparisons of the global X-ray properties of our simulations to observations because it is large, complete and presents results in which the cores of systems are not excised (a common procedure in other catalogues performed to reduce scatter in scaling relations for cosmological studies).

\subsection{Classification of Morphological States}\label{analysis-morphology}

We seek to determine how our evolving merger systems compare with recent \Chandra\ and \XMM\ observations 
of the ICM: at which points they would manifest as compact or
extended core systems, when they would appear to have been significantly
affected by radiative cooling based on their entropy and projected
temperature profiles, and when they would exhibit evidence of merger
activity under typical observational circumstances.  To minimize
biases in our comparison to observations, we have developed a
classification scheme based on a set of easily observable properties
of the system which can be readily related to its physical state.

It has been noted that projection effects can have a significant
influence on the surface brightness structure and projected
temperatures of a merging system \citep{RT02}.  For this reason, we have 
considered three orthogonal projections in our study: two in the plane of the orbit along the line initially connecting
the systems (x) and in the direction of initial tangential velocity
(y) and one normal to the plane of the orbit (z). However, given the rarity of 
observing systems in a configuration corresponding to our x-projections, we shall 
focus only on the y and z-projections in this paper.  The only significant effect of this omission  
is to exclude those rare and peculiar states produced when the systems are
separate and still relaxed but appear as one system in projection.

\subsubsection{surface brightness}

We first classify our systems according to the shape of their
projected surface brightness profiles.  To do so, we utilize the
simple 3-parameter $\beta$-model commonly employed in analysis of
observations.  This model characterizes the azimuthally averaged
surface brightness profile of a system at a projected
radius$^{7}$\footnotetext[7]{Throughout this paper, we adopt the peak
  of the projected bolometric X-ray surface brightness as the center
  of the system.} $r_p$ by the equation

\begin{center}
\begin{equation}\label{eqn-Sigma}
\Sigma_x(r_p)=\frac{\Sigma_o}{{\left( 1+ {\left( r_p/r_c \right)}^{2} \right)}^{(6\beta-1)/2}}
\end{equation}
\end{center}

\noindent where $r_c$ is the profile's core radius, $\beta$
characterizes its asymptotic logarithmic slope at large radius and
$\Sigma_o$ its normalization.  We exclude projected radii less than
$30$\kpc\ from this analysis and have verified
that varying this radius from $20$ to $50$\kpc\ has a minimal effect
and does not influence the conclusions of our study.

Although more sophisticated parameterizations of the X-ray surface brightness
profiles of clusters have been developed \citep[\eg][]{Vikhlininetal06, Mahdavietal07}, we utilize this one
to take advantage of the large statistical samples of $\beta$-model
fits presently available.

To separate ``compact'' cores from ``extended'' cores we shall use a cut of $100$\kpc.  This corresponds approximately to the unpopulated region of the observed $r_c$ distribution reported by \citet{OtaMitsuda04} separating observed compact and extended core systems.

\subsubsection{central entropy and temperature gradient}

To judge the relevance of radiative cooling for our evolving cluster 
cores we will use the central entropy of our systems.  We do this by 
computing radially binned entropy profiles, denoted as $S(r)$, and 
interpolate from them to obtain ``central'' values at $40$\kpc\ (denoted as $S_{40}$).  We use 
entropy for this purpose rather than the system's projected
temperature because it is numerically robust
in our simulations and unambiguously tied to the
relative importance of heating and cooling for both our observed and
simulated systems. Furthermore, as pointed out previously by several
authors
\citep[\eg][]{Voitetal02,Voitetal03,KaiserandBinney03,McCarthyetal07},
variations in the core properties of X-ray clusters are most insightfully
understood in terms of this quantity.

To separate ``cool'' core systems from ``warm'' core systems, we shall use a cut in $S_{40}$ of $50$\keVcmsq\ and refer to systems with $S_{40}<50$\keVcmsq\ as ``cool core'' systems and systems with $S_{40}>50$\keVcmsq\ as ``warm core'' systems.  This choice is motivated by an examination of the entropy profiles of the \citet{Sandersonetal06} catalogue (these will be presented in a future publication); central entropies lower than this value tend to occur in cool core systems while higher central entropies tend to occur in warm cores.  Furthermore, recent examinations of BCGs in galaxy clusters find a break at this entropy level such that those in clusters with $S_{40} \lesssim 50$\keVcmsq\ show
evidence of star formation while those in systems with warmer cores do not \citep{Bildfelletal08, Raffertyetal08}.  For reference, the cooling time of material with $S=50$\keVcmsq\ is $\sim2$\Gyrs\ for our systems. 

However, temperature is an intuitively appealing quantity and can be more directly
compared to observations.  For this reason we have also measured
the shape of our systems' evolving projected central temperature
profiles.
\Chandra~ and \XMM~ observations have shown that projected cluster temperature
profiles have a universal shape outside of the projected radius $0.1R_{200}$; a broad peak between $0.1R_{200}$
and $0.2R_{200}$ and a continuous decline outside of $0.2R_{200}$
\citep{Markevitchetal98,Sandersonetal06,Prattetal06}.  For projected radii within $0.1R_{200}$, a great
deal of system-to-system variation in temperature profiles is seen.
This radius roughly corresponds to the cooling radius of compact core
clusters and it is within this region that cool core systems exhibit
relatively low temperatures \citep{Piffarettietal04}. 

Motivated by these observations, we have chosen to characterize the
apparent temperature structure of the cores of our merging systems by
measuring the strength of their central temperature gradients with the
quantity 

\begin{center}
\begin{equation}\label{eqn-DeltaT}
\Delta_T=\frac{T_{outer}-T_{inner}}{T_{outer}}
\end{equation}
\end{center}

\noindent Here, $T_{inner}$ is the spectrally fit temperature (see
Poole07 for details regarding our method of simulating observed
cluster temperatures) of material in projected radii
from $30$\kpc\ to $0.1R_{200}$ ($0$ to $0.1R_{200}$ for the observations; this 
represents a negligible difference however) and $T_{outer}$ the spectrally
fit temperature for material in the range of projected radii
$0.1R_{200}$ to $0.2R_{200}$.  This quantity is thus a
measure of the temperature gradient within the core region, where
system-to-system variations are observed and cool core systems are
seen to have low temperatures.  Defined in this way, $\Delta_T$ is
positive for systems with core temperatures cooler than isothermal and
negative for systems with core temperatures warmer than isothermal.

The entropy cut we have chosen ($S_{40}=50$\keVcmsq) for dividing cool 
cores from warm cores roughly translates to a cut in central temperature gradients of \DT$=0.05$ for relaxed systems. 

\subsubsection{apparent dynamical state}

Each projection for each system is also classified into instances
where it would exhibit no substructure and regular isophots to an
observer, significant substructure, or so much substructure as to
render the system indescribable by a $\beta$-model (generally due to
the close proximity of the secondary or various other brief but extreme
distortions).  We refer to these situations as undisturbed, disturbed
and strongly disturbed respectively.  Classification is performed
through visual inspection of simulated $50$ks \Chandra~ observations
for each output (with the system at $z=0.1$), in each projection.
Our simulated observations were produced following the procedure detailed in Poole06.

Since instances of strongly disturbed morphology can not be assigned meaningful core radii, they are omitted from this morphological discussion.  Such instances are rare and their exclusion does not represent a significant omission, although it is important to keep in mind that this exceptional class does manifest occasionally at early stages of the interaction. 

\begin{figure*}
\begin{minipage}{175mm}
\begin{center}
\leavevmode \hspace*{-1cm} \epsfysize=19cm \epsfbox{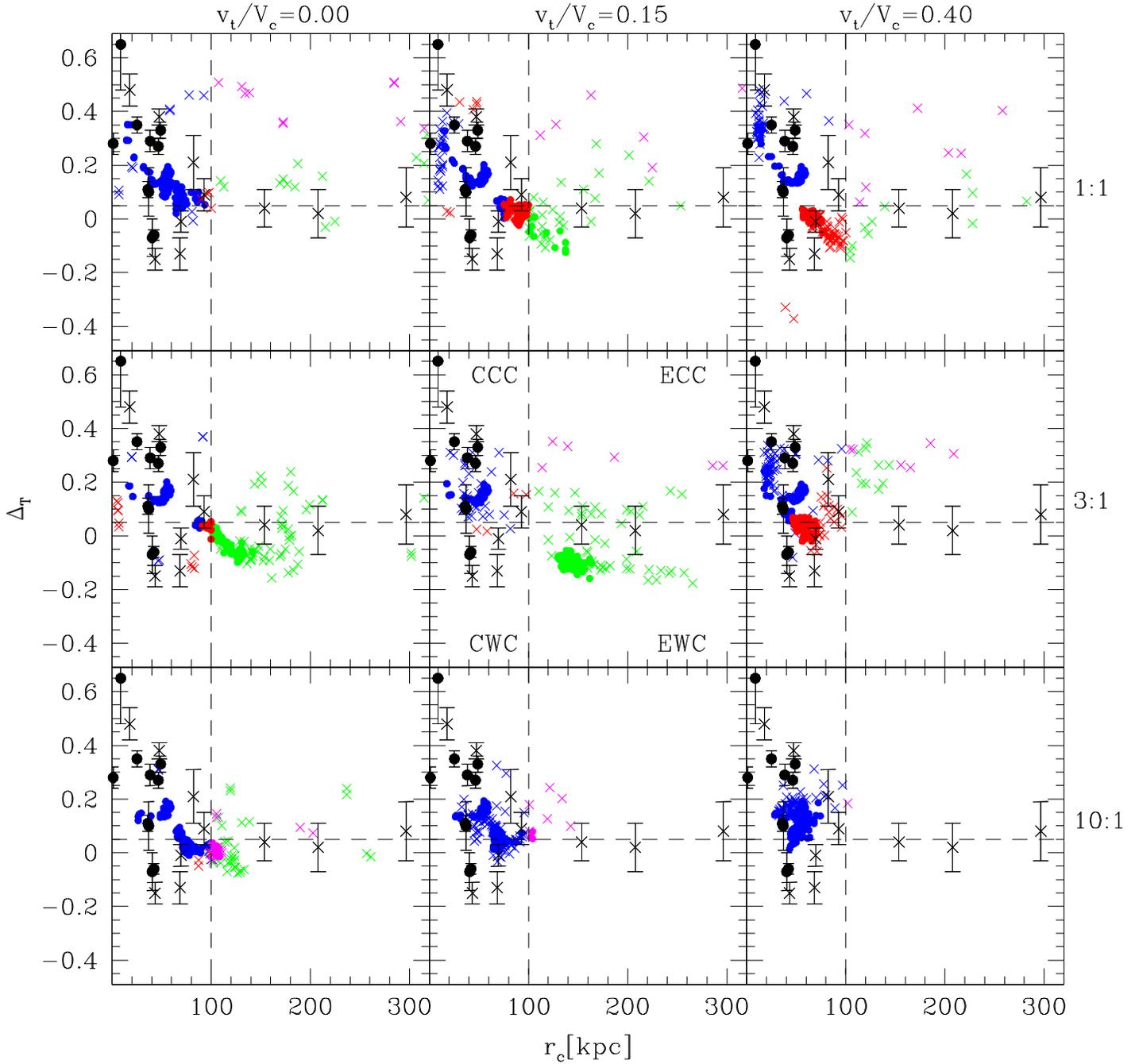}
\caption[Central temperature gradient (\DT) plotted against core radius (\rc)]{Central temperature gradient (\DT) plotted against core radius (\rc) for each output of our nine simulations in y and z-projections.  The cuts in core radius we use to separate compact-core systems ($r_c<100$\kpc) from extended-core systems ($r_c>100$\kpc) as well as the cut in \DT\ which roughly separates cool core ($S_{40}<50$\keVcmsq) from warm core ($S_{40}>50$\keVcmsq) systems are shown with dashed lines.  Compact core systems with low/high central entropies (\ie\ CCC and CWC systems) are labeled in blue and red respectively while extended core systems with low/high central entropies (\ie\ ECC and EWC systems) are in magenta and green respectively.  The quadrants of this plane corresponding to each morphology are labelled in the central pannel.  Each population is further subdivided into systems showing obvious evidence of substructure (crosses) and systems with regular isophots (solid circles).  Data from the statistically representative catalogue of \citet{Sandersonetal06} is shown in black with error bars (closed circles and crosses indicate systems noted in their study to be relaxed or visibly disturbed respectively).  Black text around the boundary indicate the mass ratio and $v_t/V_c$ of the simulations presented in each panel.}
\label{fig-delta_rc}
\end{center}
\end{minipage}
\end{figure*}

\begin{figure*}
\begin{minipage}{175mm}
\begin{center}
\leavevmode \hspace*{-1cm} \epsfysize=19cm \epsfbox{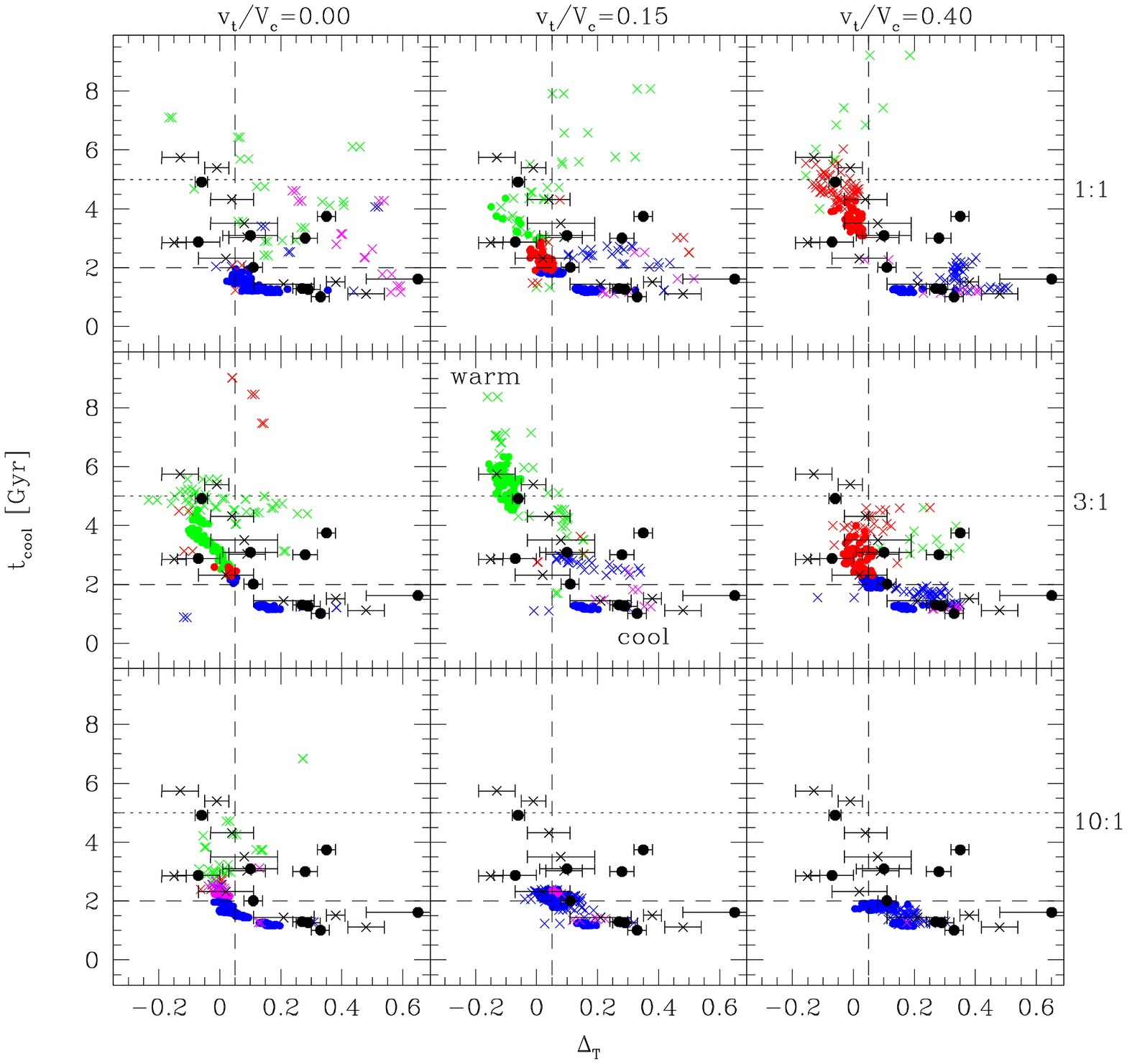}
\caption[Central cooling time plotted against central temperature decrement]{Central cooling time ($t_{cool}$) plotted against central temperature decrement (\DT) for each output of our nine simulations in y and z-projections (following the same format as Fig. \ref{fig-delta_rc}).  The cuts in \DT\ and $t_{cool}$ which roughly separate cool core systems from warm core systems are shown with dashed lines (the quadrants corresponding to warm and cool morphologies are labelled in the central panel) while the $t_{cool}=5$\Gyr\ cut used by \citet{Sandersonetal06} to divide short cooling time systems from long cooling time systems is shown with dotted lines.  Compact core systems with low ($S_{40}<50$\keVcmsq) and high ($S_{40}>50$\keVcmsq) central entropies (\ie\ CCC and CWC systems) are labeled in blue and red respectively while extended core systems with low/high central entropies (\ie\ ECC and EWC systems) are in magenta and green respectively.  Each population is further subdivided into systems showing obvious evidence of substructure (crosses) and systems with regular isophots (solid circles).  Data from the statistically representative catalogue of \citet{Sandersonetal06} is shown in black with error bars (closed circles and crosses indicate systems noted in their study to be relaxed or visibly disturbed respectively).  Black text around the boundary indicate the mass ratio and $v_t/V_c$ of the simulations presented in each panel.}
\label{fig-tcool_DeltaT}
\end{center}
\end{minipage}
\end{figure*}

\subsection{Three Dominant Morphologies}\label{analysis-dominant}

In Fig. \ref{fig-delta_rc} we present a \DTrc\ plot for our nine simulations.   Projections in y and z are displayed for each of the 120 outputs (spanning $12$\Gyr\ in $100$\Myr\ intervals) of our simulations with instances of undisturbed and disturbed morphology indicated by filled circles or crosses respectively.  Our cuts in \rc\ and \Sforty\ separate four potential populations which we identify in this plot using colours: compact cool cores (CCCs) are shown in blue, compact warm cores (CWCs) are in red, extended cool cores are in magenta and extended warm cores (EWCs) are green.  We also plot the catalogue of \citet{Sandersonetal06} in black for comparison.

Fig. \ref{fig-delta_rc}  shows that very few instances of extended cool cores occur.  Looking carefully at those circumstances which have been identified as an extended cool core, we find that in many cases it is a result of one of two effects.  In some instances when the system appears disturbed, the passage of a merging core near to the primary core (either during first or second pericentric passage) can enlarge estimates of $r_c$ while the primary core remains sufficiently undisturbed to be cool.  The system is very obviously disturbed at these times but not so much that a $\beta$- model can not be reasonably fit to its surface brightness.  The only case in which significant numbers of relaxed extended cool core states occur is the 10:1 head-on merger.  The system has relaxed to a state with core radii only slightly larger and central entropies only slightly less than our morphology cuts in this case.

With exceptional instances such as these aside, we see that throughout the duration of a merger an observer would classify the system as existing predominantly in one of three states: compact systems with cool or warm cores (CCC or CWC) or extended warm cores (EWC).  In each case, the system may or may not also possess obvious substructure and/or disturbed isophots, meaning that members of each class could be found which appear to be disturbed or undisturbed.  Including the approximate cut in \DT\ which results from our cut in $S_{40}$, we see that these three populations (especially the undisturbed set) map well to three separate regions in the \DT$-r_c$ plane.  Encouragingly, we also find good agreement between the regions of this plane occupied by our simulations and by the observations.  In Section \ref{analysis-states} we will examine the evolution of our systems through this plane and in Section \ref{analysis-structure}, examine the processes principally responsible for setting a system's morphology. 

It has been noted that clusters with warm cores can have cooling times less than their median formation age \citep{Donahueetal05,Sandersonetal06}.  In Fig. \ref{fig-tcool_DeltaT} we plot the central cooling time (measured at $40$\kpc) against $\Delta_T$ from the \citet{Sandersonetal06} catalogue and for each of our simulations.  We can see from this figure that our simulations capture the relationship between cooling time, core temperature and core size observed by \citet{Sandersonetal06}:  warm cores can have short cooling times if they are compact.  This is a straightforward consequence of the strong dependence of cooling on density and its weak dependence on temperature.

Although compact cores are not necessarily cool they nevertheless have
short cooling times (even when disturbed), preserving the importance
of radiative cooling in establishing their structure and evolution
during and after a merger.  Furthermore, although extended core
morphologies generally (but not necessarily) have longer cooling times
than compact core systems, only one of the cases we have studied produces
states with cooling times greater than $5$\Gyrs\ for any significant length of time:
our 3:1 $v_t/V_c=0.15$ case.

\subsection{Temporal Evolution of Morphology}\label{analysis-states}

To see how the morphologies of our simulations progress through these
three states, we present the temporal evolution of \rc~ and \DT~ in
Figs. \ref{fig-rc_t} and \ref{fig-Delta_t}.  In both cases, a black
curve traces the system's evolution as seen in the $z$-projection with
coloured points plotted (following the format of
Fig. \ref{fig-delta_rc}) to indicate each interval's morphological
state.  These curves change very little when viewed in $y$-projections
and differ only at early times in the $x$-projection due to the
apparently close proximity of the two cluster cores prior to
interaction (the system can appear as a ECC system in x-projections prior to \tclosest).  
To clarify these plots, we show only the $z$-projection. We also 
note \tclosest\ and \trelax\ for each simulation to indicate the interval over 
which the system will tend to look disturbed.

\subsubsection{core size}\label{analysis-states-core_size}

In Fig. \ref{fig-rc_t} we consider the evolution of $r_c$ for our simulations.
We see that in all cases the system starts with a compact core (by
construction) and experiences a period of substantially increased core
size (often $r_c>300$\kpc) during first pericentric passage.
For the head-on mergers, we see that in the 1:1 and 3:1
cases this lasts (with varying amplitude) until roughly \taccrete\
while in the 10:1 case it lasts for only $\sim$$0.5$\Gyr.  These
elevated values of $r_c$ are produced by the explosive expansion of
the core after the secondary's impact. 

In off-axis cases, these elevated values are primarily a product of
the close proximity of the cluster cores during pericentric passage.
In these cases there is an interval between \tclosest~ and \taccrete~
during which the system returns to a compact core state.  This
indicates that the initial interaction of the cores fails to disrupt
the primary system to an extended core state in any of our off-axis
mergers. 

Following \taccrete\ when the remainder of the secondary core accretes
to the remnant core, many of our merger remnants are in, or nearly in,
extended core states.  They subsequently evolve back towards their
initial core size (rapidly in some cases) with only the 3:1 head-on
and 1:1 and 3:1 $v_t/V_c=0.15$ cases spending prolonged periods of time in
extended core states.  Of these, only the 3:1
$v_t/V_c=0.15$ case fails to formally recover a compact core by the end of the simulation.  The remnant in this case reaches an extended core state with $r_c\sim140$\kpc\ following
\taccrete~ and evolves very little afterward.  

In Section \ref{analysis-structure} we will discuss the physical processes which shape the
diversity in remnant core states. 

\subsubsection{central temperature gradients}\label{analysis-states-temperature}

\begin{figure*}
\begin{minipage}{175mm}
\begin{center}
\leavevmode \epsfysize=8.4cm \epsfbox{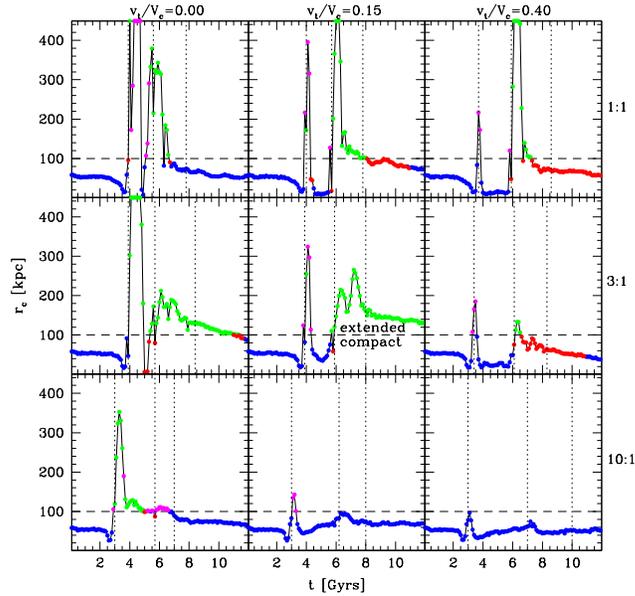}
\caption[Temporal evolution of core radius]{The temporal evolution of
  core radius (\rc) for our simulations.  Black curves illustrate
  $r_c(t)$ and coloured points indicate the morphology of the system measured in the
  $z$-projection.  Compact core systems with low ($S_{40}<50$\keVcmsq) and high ($S_{40}>50$\keVcmsq) central entropies (\ie\ CCC and CWC systems) are labeled in blue and red respectively while extended core systems with low/high central entropies (\ie\ ECC and EWC systems) are in magenta and green respectively.  The cut in $r_c$ used to separate compact
  core systems from extended core systems is shown with dashed lines (see labels in central panel).
  Vertical dotted lines indicate (from left to right) \tclosest,
  \taccrete\ and \trelax\ for each case.  The interval between \tclosest\ and \trelax\ indicates the time during which the
  system will show signs of being disturbed.  Black text around the boundary indicate the mass ratio and $v_t/V_c$ of the simulations presented in each
  panel.} 
\label{fig-rc_t}
\end{center}
\end{minipage}
\end{figure*}

\begin{figure*}
\begin{minipage}{175mm}
\begin{center}
\leavevmode \epsfysize=8.4cm \epsfbox{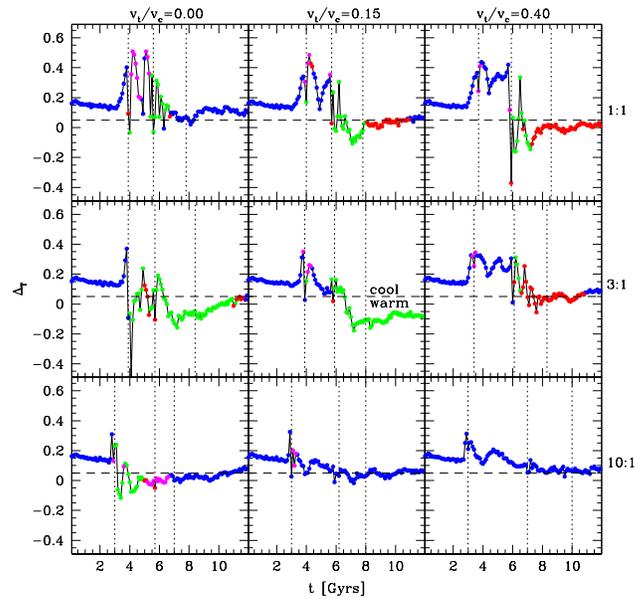}
\caption[Temporal evolution of central temperature gradient]{The
  temporal evolution of the central temperature gradient (\DT)
  following the same format as Fig. \ref{fig-rc_t}.  Black curves
  illustrate $\Delta_T(t)$ and coloured points indicate the morphology
  of the system
  measured in the $z$-projection.   Compact core systems with low ($S_{40}<50$\keVcmsq) and high ($S_{40}>50$\keVcmsq) central entropies (\ie\ CCC and CWC systems) are labeled in blue and red respectively while extended core systems with low/high central entropies (\ie\ ECC and EWC systems) are in magenta and green respectively.  The \DT$=0.05$ line roughly separating cool core systems from
  warm core systems (classified by entropy) is shown with dashed lines (see labels in central panel).  Vertical dotted lines
  indicate (from left to right) \tclosest, \taccrete\ and
  \trelax\ for each case.  The interval between \tclosest\ and \trelax\ indicates the time during which the
  system will show signs of being disturbed.  Black text around the boundary indicate the
  mass ratio and $v_t/V_c$ of the simulations presented in each panel.} 
\label{fig-Delta_t}
\end{center}
\end{minipage}
\end{figure*}

In Fig.\ref{fig-Delta_t} we plot the evolution of $\Delta_T$ for our
simulations following the format of Fig. \ref{fig-rc_t}.  In all
cases our systems begin with strong positive central temperature
gradients (by construction) which remains intact until $t_o$, modulo a very slow decline due to
the numerical effects discussed in Section \ref{numerics-cores}. 


Considering the head-on mergers first, we see that in the rare 1:1
head-on case,  $\Delta_T$ remains above $0$ at nearly all times.
In the more common 3:1 and 10:1 head-on cases, the system briefly develops a
negative central temperature gradient after \tclosest\ and oscillates
with a warm core two or three times on dynamical timescales. 

In off-axis cases, the primary core's positive central temperature
gradient survives until \taccrete.  This indicates that the first
interaction of the cores in off-axis cases does not heat the primary
core to a warm state.  At (or shortly following) \taccrete\ however, all of our
1:1 and 3:1 mergers experience a rapid decline in \DT.  In the 10:1 cases there is
little evolution in \DT\ following \taccrete\ due to the substantial disruption of the secondary system
prior to second pericentric passage in these cases.

Hence, we find that cool cores resiliently maintain their
integrity until a merging system's core makes direct impact, either at
\tclosest\ in head-on cases or during second pericentric passage at
\taccrete\ in 1:1 and 3:1 off-axis cases.  Extended cores are not generically
produced (except as very short lived transient states during which the
system would appear visibly disturbed) and the cores generally recover cool states once relaxed
(the most significant exceptions are the 3:1 head-on and $v_t/V_c=0.15$ cases).  

We explore the physical processes which shape the diversity of remnant core states in Section \ref{analysis-structure}.

\subsection{The recovery of compact cool cores}\label{analysis-structure}

As we can see from both Figs. \ref{fig-rc_t} and \ref{fig-Delta_t}, a
pattern emerges for the durations of EWC states (illustrated in green)
produced in our mergers: near-axis (head-on and $v_t/V_c=0.15$) and
intermediate mass ratios (3:1) are the most efficient cases for
producing this morphology.  To understand this, we now seek to
explain two things: how the initial conditions of the merger (namely, the mass
ratio and orbits of the clusters) dictate the final remnant state
which emerge at \trelax\ and what properties of the remnant core's
structure principally govern the recovery of CCC morphologies.   

In essence, this reduces to an examination of the final balance
between shock heating and radiative cooling for material destined to
end-up in the core of the remnant.  Thus, the pattern of sustained EWC
morphologies seen in Figs. \ref{fig-rc_t} and \ref{fig-Delta_t} can be
understood as the result of a ``Goldie Locks'' scenario where the
circumstances necessary for effective core heating are optimized while
the formation of dense structures capable of offsetting heating
through radiative cooling are prevented. 

These competing processes are both sensitive to the evolving dynamical
state of the initial systems' cores and the relative densities and
mach numbers of gaseous material forced to interact during the event.
We find that a great variety of processes are involved in establishing
this balance, with the effects of both shock heating and radiative
cooling being a sensitive function of the mass ratio and initial
orbits of the interacting clusters. 

For the remainder of this section we shall examine this variety.
However, as discussed in Section \ref{numerics-statistics}, our sparse
sampling of $M_p:M_s$ and $v_t/V_c$ prevents us from systematically
studying many of these processes in a quantitative way.  We will thus
restrict ourselves to a heuristic discussion in what follows. 

\begin{figure*}
\begin{minipage}{175mm}
\begin{center}
\leavevmode \epsfysize=10cm \epsfbox{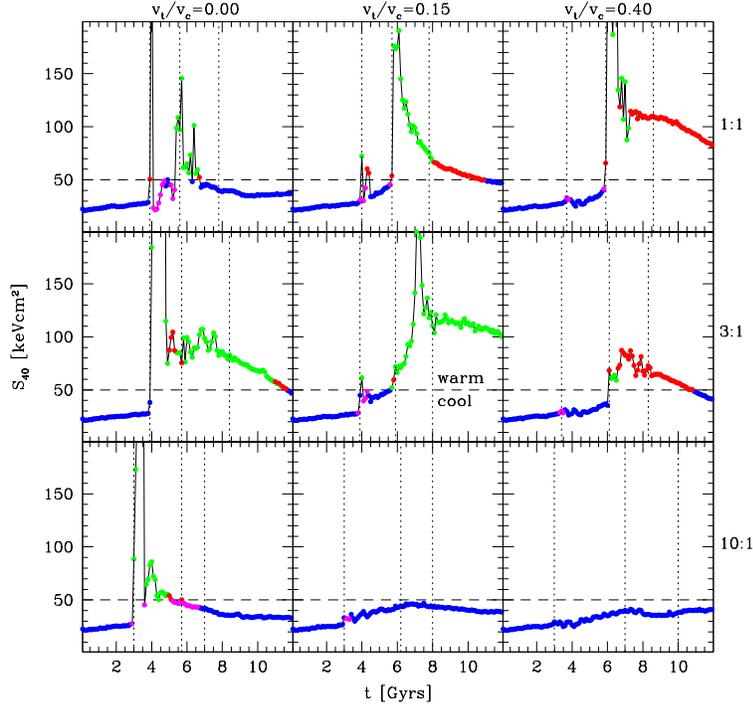}
\caption[Evolution of central entropy]{The
  temporal evolution of the central entropy (measured at $40$\kpc) 
  following the same format as Fig. \ref{fig-rc_t}.  Black curves
  illustrate $S_{40}(t)$ and coloured points indicate the morphology
  of the system measured in the $z$-projection.   Compact core systems with low ($S_{40}<50$\keVcmsq) and high ($S_{40}>50$\keVcmsq) central entropies (\ie\ CCC and CWC systems) are labeled in blue and red respectively while extended core systems with low/high central entropies (\ie\ ECC and EWC systems) are in magenta and green respectively.  The cut in $S_{40}$ used in our morphological classification to separate cool core systems from
  warm core systems is shown with dashed lines (see labels in central panel).  Vertical dotted lines
  indicate (from left to right) \tclosest, \taccrete\ and
  \trelax\ for each case.  The interval between \tclosest\ and \trelax\ indicates the time during which the
  system will show signs of being disturbed.  Black text around the boundary indicate the
  mass ratio and $v_t/V_c$ of the simulations presented in each panel.} 
\label{fig-entropy_t}
\end{center}
\end{minipage}
\end{figure*}

\subsubsection{entropy and the formation of extended cores}\label{analysis-entropy}

The temperature structure of a relaxed cluster's ICM is primarily set
by its gravitational potential.  Because of this and the strong
$\rho_g^2$ dependence of the ICM's X-ray emissivity, variations in the
central X-ray surface brightnesses of clusters (and hence, variations
in $r_c$) are primarily a product of variations in central gas
densities.  Because it is directly tied to the mass accretion history of the system and the
relative balance between heating (from accretion shocks and feedback) and radiative cooling, 
these density variations are most straightforwardly interpreted in terms of variations in central (or minimum) gas entropies.

In Fig. \ref{fig-entropy_t} we present the temporal evolution of the
central entropies of our simulations where we see a good
correlation between $S_{40}(t)$ and $r_c(t)$ following \trelax.  Hence, the 
surface brightness core radii of our merger remnants are tightly related to their 
central entropy.  We now address the question: what processes set this entropy?   

As in the plots of $r_c(t)$ and $\Delta_T(t)$, we observe important
differences in $S_{40}(t)$ between head-on and off-axis cases.
In the head-on cases, one very dramatic increase in $S_{40}$ occurs at
\tclosest\ followed by several small and erratic oscillations which
end $\sim1$\Gyr\ before \trelax.  The details of this evolution depend
quite sensitively on the mass ratio of the event.  Examining the
dynamical evolution of these mergers in the spatial distribution of
their surface brightness, temperature and entropy (Figs. 1 to 9 in
Poole06) we can see why this is so.  In the 3:1 and 10:1 cases, the
secondary core fully penetrates the primary core triggering explosive shocks which 
heat the entire primary core from the inside out.  In the
1:1 case, the merging cores inelastically collide and remain in very
dense structures following \tclosest.  This has the dual effect of
reducing the efficiency of entropy production through shocks and
increasing the rate of radiative cooling.  Together, these effects
lead to a counter intuitive result: the 3:1 and 10:1 head-on mergers
ultimately heat the primary core more than the 1:1 head-on merger. 

In off-axis cases, there are slight and brief increases in central
entropy at \tclosest\ in some cases, but there is generally very
little sustained increase in $S_{40}$ until \taccrete.  This is
because the initial interaction of the systems has very little effect
on the primary core and is consistent with what we noted in Section
\ref{analysis-states}: the primary core retains its initial CCC
morphology until second pericentric passage in off-axis cases.  In the
1:1 cases, the secondary core remains significantly intact until
\taccrete\ and we thus see a significant jump in entropy.  In other words: the evolution of the 1:1 $v_t/V_c=0.15$ case following \taccrete\
is very similar to the 3:1 head-on case following \tclosest.  In the 10:1
off-axis cases, the secondary core is so significantly disrupted by
\taccrete\ that essentially no evolution in central entropy is seen
throughout the entire simulation.  

In the 3:1 off-axis cases, an evolution distinctly different from the 1:1 and 10:1 
off-axis cases is observed.   In Poole06 we used our 3:1 $v_t/V_c=0.15$ case to illustrate how an
off-axis merger of a cold-core system can generate low-entropy streams
of material which accrete to the remnant core with high velocities.
Fig. \ref{fig-entropy_t} illustrates that the accretion of these
streams can have a significant impact on the central entropy of
off-axis merger remnants.  In both off-axis 3:1 cases we see a very
low amplitude jump at \taccrete\ when the surviving portion of the
secondary core returns for its second pericentric passage.  It is not
until significantly after \taccrete\ that peak levels of $S_{40}$ are
reached.  These peaks occur roughly when the streams formed from the
secondary cores are accreting most significantly onto the remnant. 

The heating effects of an accreting stream depend sensitively on its
total mass, density, and mach number as well as the state of the
remnant core onto which it falls.  The ram pressure and tidal forces
experienced by the cores of interacting systems, which are responsible
for stripping the secondary, forming the stream and disturbing the
primary core, depend very sensitively on the initial mass ratio and
impact parameter of the merger.  The stream's mach number during
accretion also depends on the amount of dynamical friction experienced
by the secondary system during its orbit, and thus, the mass ratio of
the interacting systems.  Hence, the heating effects of stream
accretion are very sensitive to the initial mass ratio and orbits of
the interacting clusters. 

Thus, the central entropy of the remnant, which is a very sensitive
function of the mass ratio and initial orbit of the merger, plays a
key role in setting its core size.  In the cases
we have studied, the only long lived extended core remnant results from a slightly
off-axis 3:1 merger for which stream heating is the dominant mode of
core entropy production. 

\subsubsection{cooling timescales and cool core recovery}\label{analysis-timescales}

Radiative emission can reduce the temperature of material, but must it do so?  
Imagine a disturbed parcel of low-entropy gas radiatively cooling as it falls to the center of the system.  As it cools, its
entropy will fall on its cooling timescale, $t_{cool}\propto
S^{1.5}T^{-1}$.  However, because of its dynamically disturbed state, 
its temperature can drop only if the cooling timescale is sufficiently
short with respect to its local dynamical time, $t_{dyn}\propto (G
\Delta \rho_c)^{-0.5}$.  If $t_{cool}\gg t_{dyn}$, temperature increases from  
compression dominate over radiative losses.  Under these 
circumstances, the temperature can only fall once the material reaches a quasi-static state.
It will then do so on timescales governed by the cooling time.  

At all times following \taccrete\ and for all the cases we have studied, we find that
$t_{cool}\gg t_{dyn}$ within $r_c$ (by approximately an order of magnitude) for our disturbed but relaxing remnants.  As a
result, cooling does not lower our remnant cores' temperatures until
they have returned to quasi-stable states of hydrostatic equilibrium and a substantial fraction of a 
cooling time has passed.  In Poole06 we examined how a system returns to hydrostatic equilibrium
following a merger.  We showed that our remnants generally return to
within $10$\% of hydrostatic equilibrium at $R_{500}$ $0.5-1.0$\Gyrs\ following 
the system's apparent relaxation at \trelax.  Examining this behavior at $R_{2500}$ 
(roughly $r_c$) we find that the system returns to within $10$\% of hydrostatic equilibrium 
a little sooner and generally within $0.5$\Gyrs\ of when it looks relaxed at \trelax.  Hence, cooling 
generally will not have a significant effect on central remnant temperatures until a time $t_{cool}$ 
(\ie\ $\sim2-5$\Gyrs) after appearing relaxed at \trelax.  This accounts for the occurrence of compact 
warm core (CWC) morphologies as well as the timing of the late increases in \DT\ occurring in several of our 1:1 and 3:1 merger 
remnants (whose central cooling times are short).

To conclude, we find that following a disturbance from a merger, the initially dense cores
of CCC systems generally recover much faster (roughly a few dynamical times following $t_o$) 
than their cool temperatures (roughly $t_{cool}$ afterward).  The existence 
of compact warm core (CWC) systems observed by \Chandra\ and \XMM\ \citep{Sandersonetal06,Prattetal06} can 
thus be understood naturally within the framework of hierarchical structure formation without appealing to other heating 
sources such as AGN.  Their existence is merely a natural consequence of hierarchical structure formation 
and the fact that $t_{cool}\gg t_{dyn}$ in cluster cores.

\subsection{Morphological segregation in scaling relations}\label{analysis-segregation}

Scaling relations between mass, X-ray temperature and luminosity are
powerful tools for studying the processes that shaped the structure of
the ICM.  It has been known for some time
\citep{Fabianetal94,Markevitch98} that a significant fraction of the
scatter in these relations is due to variations in the core properties
of clusters.  This has been supported by recent theoretical
\citep[][Poole07, McCarthy et al. 2007]{OHaraetal05} and observational studies
\citep{Sandersonetal06,Prattetal06}.  For this reason, and because the
generation of accurate mass functions from scaling relations has been a
significant motivator for their study, past analyzes have typically
been conducted with the central cores of systems excised to reduce scatter. 

However, it has been shown by M04 that there is interesting structure
in these relations which correlate with the core properties of
clusters.  They use the ACC catalogue of \citet{Horner_thesis} to
compare their analytic entropy injection model to published \Chandra~
and \XMM~ temperature profiles, classifying them by the presence or
absence of central temperature gradients.  They further separate these
systems into classes showing evidence of significant or little
substructure.  Despite the small number of published \Chandra~ and
\XMM~ temperature profiles available to them at the time, they found
that these classes clearly (but not necessarily distinctly) segregate
in both the \LT~ and \betarc~ planes. 

In Poole07 we examined the evolving global properties (including
X-ray temperature and luminosity) of our simulations, as well as the
scaling relations generated from them.  Having quantified the evolving
morphologies of our cluster cores, we now examine how the three
morphology classes we have identified segregate in these planes.  In
Poole07 we found that the evolution in the $L_x-T_x$, $M_t-L_x$,
$M_t-T_x$, $SZ-L_x$ and $SZ-T_x$ planes exhibited a generic pattern
and we find the same for morphological segregation in these planes as
well.  For this reason we will only examine the $L_x-T_x$ and $\beta-r_c$ relations here. 

\subsubsection{the $L_x-T_x$ plane}\label{analysis-LT_segregation}

\begin{figure*}
\begin{minipage}{175mm}
\begin{center}
\leavevmode \hspace*{-1cm} \epsfysize=19cm \epsfbox{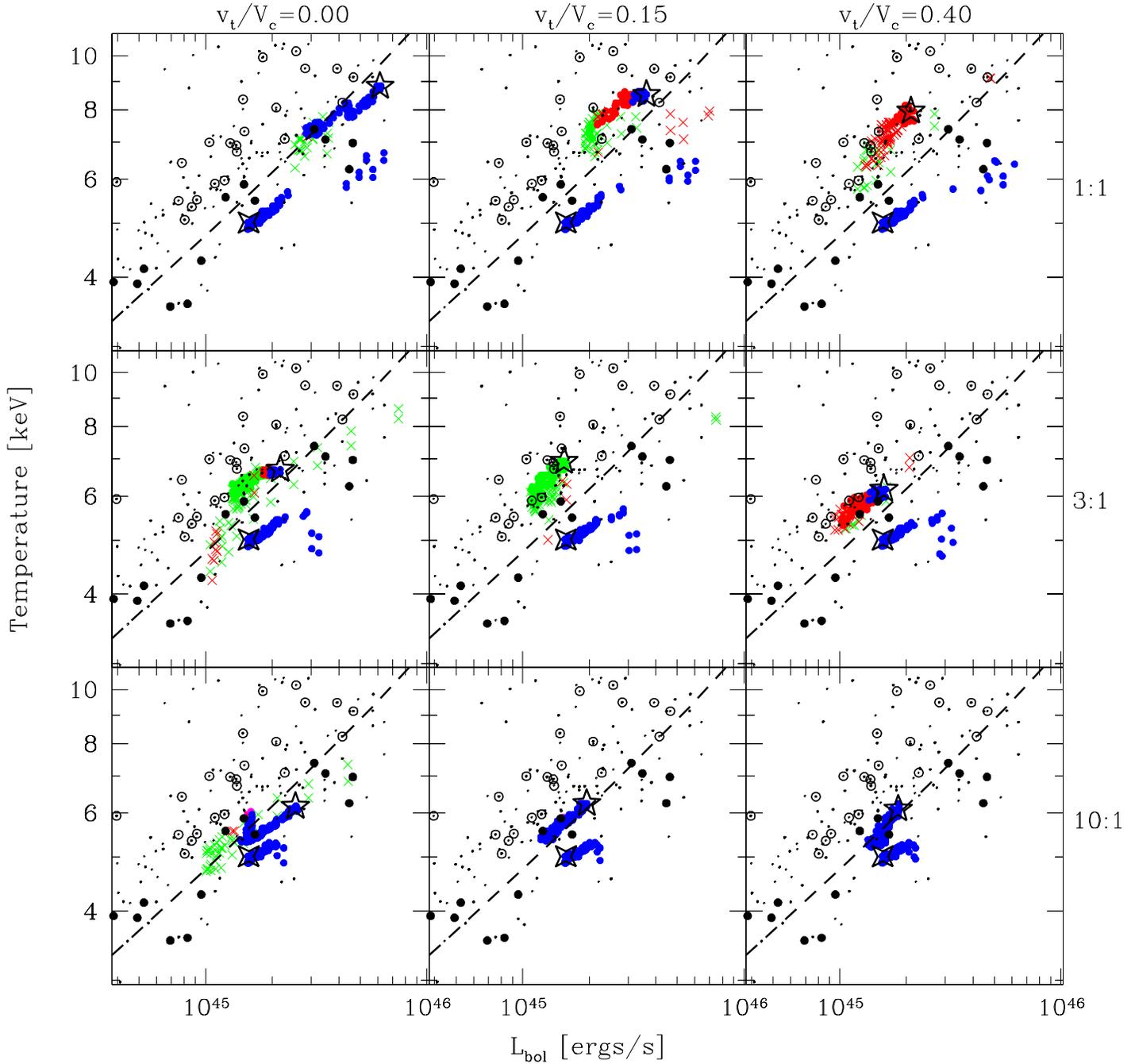}
\caption[$L_x$ against $T_x$ illustrating morphological
segregation in this plane]{A plot of $L_x$ against $T_x$ illustrating
  the morphological segregation in this plane.
  Compact core systems with low ($S_{40}<50$\keVcmsq) and high ($S_{40}>50$\keVcmsq) central entropies (\ie\ CCC and CWC systems) are labeled in blue and red respectively while extended core systems with high central entropies (\ie\ EWC systems) are in green.  Filled circles and crosses mark systems which appear relaxed
  or unrelaxed (respectively) in simulated $50$\ks\ \Chandra\
  observations.  Black dots are the
  observed catalogue of \citet{Horner_thesis} with open circles 
  marking systems with extended cores ($r_c>100$\kpc) and filled
  circles marking systems with compact cores ($r_c<100$\kpc).  The
  black dashed line is a fiducial entropy injection model with
  $S_o=150$\keVcmsq~ (see Poole07 for details).  Black text around
  the boundary indicate the mass ratio and $v_t/V_c$ depicted by each
  panel.} 
\label{fig-L_T}
\end{center}
\end{minipage}
\end{figure*}
\begin{figure*}
\begin{minipage}{175mm}
\begin{center}
\leavevmode \hspace*{-1cm} \epsfysize=19cm \epsfbox{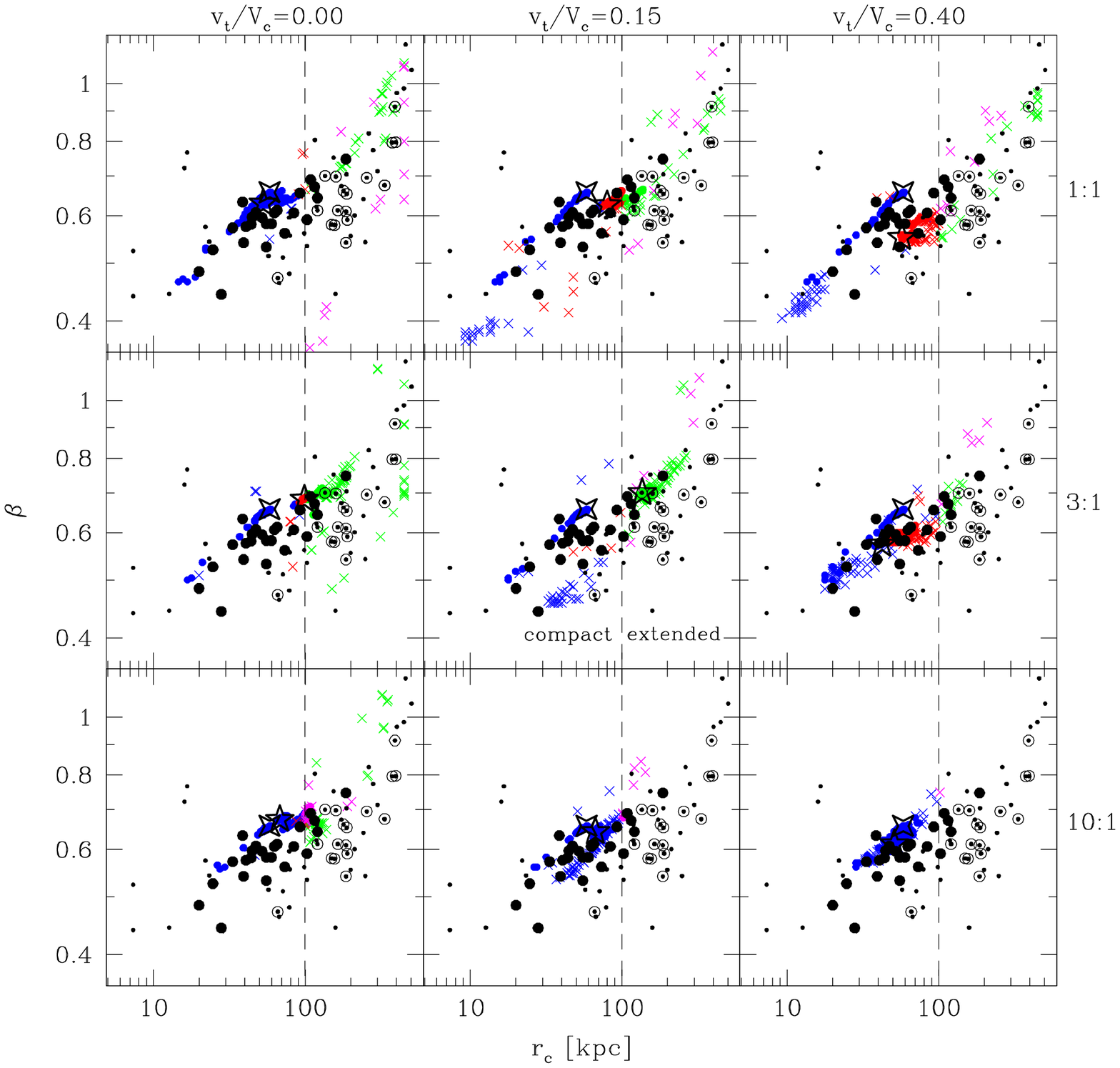}
\caption[$\beta$ against $r_c$ illustrating the morphological
segregation in this plane]{A plot of $\beta$ against $r_c$
  illustrating the segregation of morphology and of long and short cooling time systems
  in this plane.  Compact core systems with low ($S_{40}<50$\keVcmsq) and high ($S_{40}>50$\keVcmsq) central entropies (\ie\ CCC and CWC systems) are labeled in blue and red respectively while extended core systems with low/high central entropies (\ie\ ECC and EWC systems) are in magenta and green respectively.  Filled circles and crosses mark systems
  which appear relaxed or unrelaxed (respectively) in simulated
  $50$\ks~ \Chandra~ observations with the system placed at $z=0.1$.
   Black dots
  are the observed catalogue of \citet{Horner_thesis} with open
  circles marking systems with long cooling times ($t_{cool}>5$\Gyrs)
  and filled circles marking systems with short cooling times
  ($t_{cool}<5$\Gyrs).  Dashed lines mark our $r_c=100$\kpc\ cut separating compact from extended core systems (see labels in central panel).  
  Black text around the boundary indicate the
  mass ratio and $v_t/V_c$ depicted by each panel.} 
\label{fig-beta_rc}
\end{center}
\end{minipage}
\end{figure*}

In Fig. \ref{fig-L_T} we present the positions of our systems in the
$L_x-T_x$ plane, at each time in y and z-projections,
when they appear as CCC (blue), CWC (red) or EWC (green) systems.  The
initial and final states of the systems are indicated with black four
and five point stars respectively.  These are compared against the
observed catalogue of \citet[][in black]{Horner_thesis} for which we use the catalog of ROSAT 1-D $\beta$-model fits from \citet{ReiprichandBohringer02} 
to identify systems with compact cores ($r_c<100$\kpc) using filled
circles and systems with extended cores ($r_c>100$\kpc) using open
circles.   

This plot reinforces what was found in Poole07:  the low-luminosity
side of the $L_x-T_x$ scaling relation can not be significantly
populated with the mergers we have studied.  It is comforting to note
that our systems remain confined within the scatter of the
observations at all times, but instances of systems lying near the low
luminosity side of this scatter are rare and short lived. 

Comparing the regions occupied by systems which exhibit compact and
extended cores, we find a good agreement between the observations and
our simulations (with the exception of the rare 1:1 off-axis mergers,
which yield much more under-luminous compact core systems).  Furthermore, there is a very
clean divide between warm and cool cores in this plane.  We have plotted a fiducial
entropy injection model with a minimum entropy of $S_o=150$\keVcmsq\
\citep[computed following the general approach of][see Poole07 for
some additional details]{MBBPH} to roughly quantify the location of this divide.

For our simulations, instances of disturbed morphology generally lie
near or above this fiducial line while relaxed morphologies lie below it.  This is consistent
with common anecdotal observations that extended core systems
generally appear disturbed.  However, our results suggest that disturbed 
morphologies may correlate better with the presence of warm cores than  
with the presence of extended cores.  It is important to emphasize however 
that our simulations do not populate the low-luminosity side of the scatter in this plane nor do they yield relaxed extended core systems such as Abell 2034 or Abell 2631.  Trends such as these suggest that the observed scatter in the $L_x-T_x$ plane is a result of physical processes other than mergers.

We have similarly examined the other scaling relations discussed in
Poole07 and find the same qualitative behavior in all
cases. 

\subsubsection{the $\beta-r_c$ plane}\label{analysis-beta_rc_segregation}

In Fig. \ref{fig-beta_rc} we present the positions our simulations occupy in
the $\beta-r_c$ plane, at each time in y and z-projections, 
when they appear as CCC (blue), CWC (red), ECC (magenta) or EWC (green)
systems.  Instances of relaxed-looking morphology are marked with solid circles and disturbed morphology with crosses. 
The initial and final states of the systems are indicated
with black four and five point stars respectively.  These are compared
against the observed catalogue of \citet{ReiprichandBohringer02} (black dots).  
We identify systems with short cooling times ($t_{cool}<5$\Gyrs)
with large filled circles and systems with long cooling times
($t_{cool}>5$\Gyrs) with large open circles.  These cooling times are obtained by combining temperatures from \citet{Horner_thesis} with denisties derived from \citet{ReiprichandBohringer02}.

We can see from this plot that $\beta$ and $r_c$ express a clear
correlation in the observations and that our mergers 
reproduce some of its features.  In particular, our simulations account 
for the high-$\beta$ envelope of the observed scatter in this plane.  Furthermore, 
our simulations populate the narrow region occupied by the observations
at the largest values of $r_c$.  While they significantly populate this region, they 
are always visibly disturbed.  Observed systems in this region for which we have 
dynamical classifications \citep{MBBPH} are similarly all disturbed.

However, our simulations fail to produce states which last for significant lengths of time in the 
region of this plane occupied by the vast majority of observed extended core systems.  Most 
observed extended core systems have values of $\beta$ substantially less than the states 
produced by our mergers.

In Fig. \ref{fig-tcool_DeltaT} we show that our mergers rarely produce
systems with cooling times in excess of $5$\Gyrs.  This is primarily a
consequence of the cores of our systems remaining compact for most of
their evolution.  We can see in this plane that observed systems with short cooling times
are generally found along the high-$\beta$ side of the observed scatter in this plane, in 
good accord with our simulations.  Observed systems with $t_{cool}>5$\Gyrs\ lie in the 
low-$\beta$/large-$r_c$ region of the plane which our simulations fail to significantly populate.

To conclude, our simulations are able to account for the high-$\beta$ envelope occupied by 
observed systems with short cooling times as well as the tail of disturbed systems at the highest values 
of \rc.  They do not account for the bulk of observed extended core systems with long cooling times.

\begin{figure*}
\begin{minipage}{175mm}
\begin{center}
\leavevmode \epsfysize=10cm \epsfbox{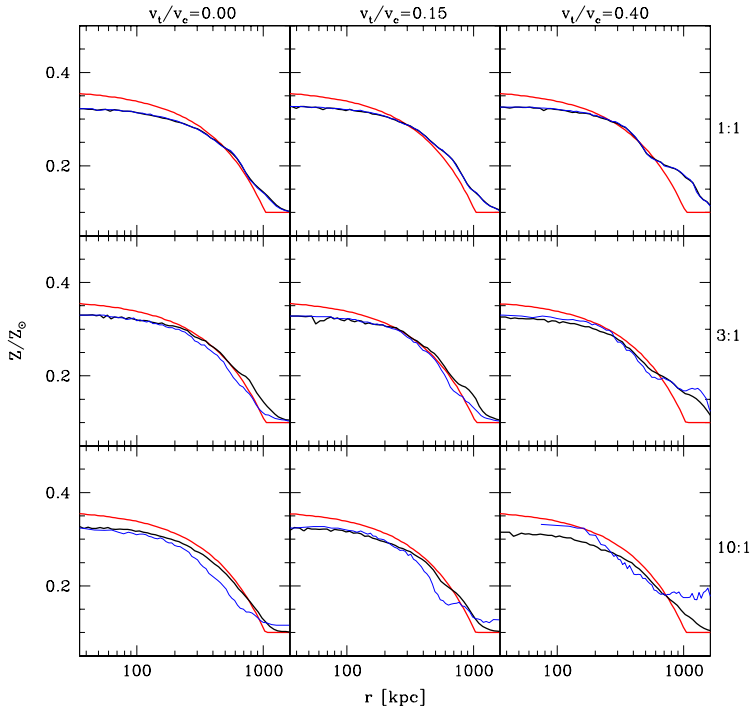}
\caption[Initial and final metallicity gradients]{A plot comparing the
  initial and final metallicity gradients of our simulations.  Initial
  profiles of the primary are shown with thick red lines and the final
  remnant profile is shown with a thick black line.  The final
  distribution of material originating in the secondary is shown with
  a thin blue line.  In all cases, the metallicity profiles of our
  systems are initialized to have the metallicity profiles observed by
  \citet{DeGrandiandMolendi02}.} 
\label{fig-Z_r}
\end{center}
\end{minipage}
\end{figure*}

\section{Metallicity gradients}\label{analysis-mixing}

It is a well established fact that extended core systems tend to lack
metallicity gradients (with mean amplitudes of $0.1-0.2Z_\odot$) while
CCC systems routinely exhibit centrally peaked metallicity gradients
within $0.2R_{180}$, often reaching central values which approach
solar metallicity \citep{DeGrandiandMolendi02}.  This, along with
anecdotal evidence suggesting that extended core systems tend to be
preferentially disturbed has been used to claim, as a result, that
mergers effectively mix the ICM. 

To examine this hypothesis, we have initialized our systems to have
the metallicity profiles presented by \citet{DeGrandiandMolendi02} and
examined the resulting metallicity profiles of our merger remnants.  We
present these results in Fig. \ref{fig-Z_r} where the initial
metallicity profiles of our primary systems are presented in blue and
the final metallicity profiles of our remnants in red. 

It is clear from this figure that mergers do not efficiently mix the
material of the initial systems.  This makes sense given the fact that
the enhanced metallicity is initially tied to the lowest entropy
material in the clusters.  This material is not preferentially heated
above the entropy of the surrounding material and thus, remains the
low entropy material constituting the remnant.  Once the system
relaxes, the low-entropy/high-metallicity material sinks to the center
of the system due to a lack of buoyant support, reestablishing a
central metallicity peak with only a slightly lower amplitude and slightly broader
distribution.

%-- DISCUSSION --
\section{Discussion}\label{sec-discussion}

In this paper, we have shown that the notions of compactness, coolness, and active cooling for cluster cores are related but not as simply as once thought.  As illustrated by the observations of \citet{Sandersonetal06}, compact cores need not be cool and can have central cooling times as long as $4$\Gyrs\ while  warm cores can have cooling times as short as 2\Gyrs.  We have shown how several aspects of these trends can be accounted for naturally in theories of hierarchical structure formation: radiative cooling naturally produces relaxed clusters with compact cool cores and significant mergers generally disturb them into warm core states with a range of core radii.  The central entropies generated by a merger event principally determine the core size of the remnant and if the resulting cooling times are short, the system will evolve back towards a CCC state after appearing relaxed.  The core dynamically relaxes much more quickly than it radiatively cools so a compact core recovers its central densities (and hence, surface brightness core size) quicker than it's cool temperatures return.  This naturally explains the population of compact warm cores observed by \Chandra.

Our mergers do not produce relaxed clusters with large-$r_c$/low-$\beta$ X-ray surface brightness profiles and long cooling times.  They do not erase metallicity gradients either.  These results imply that once established, compact cool cores are remarkably resilient to disturbances from mergers.  How then are relaxed systems with high central entropies ($S_{40}>150$\keVcmsq) and extended surface brightness cores ($r_c>150$\kpc) produced?  In Poole07 we suggested the possibility that mergers could induce AGN activity, augmenting the heating from a merger event.  Significant outbursts such as those observed by \citet{McNameraetal05} could potentially heat the cores of a disturbed system to EWC states but even MS0735.6+7421, the largest AGN outburst yet observed, still has a significant metallicity gradient.  It would appear that neither AGN outbursts nor mergers are able to sufficiently erase a metallicity gradient once it is established.

How then do we reconcile the lack of metallicity gradients in observed EWC systems with our results?  It may be that mergers yield more mixing than our simulations suggest.  SPH notoriously handles turbulence poorly which may be leading us to underestimate the degree of mixing in mergers.  On the other hand, the non-diffusive nature of SPH prevents sedimentation which would otherwise reenforce metallicity gradients.  

Alternatively, we may need to seek mechanisms by which cool cores are prevented from forming in the first place.  One means may be through variations in mass accretion histories.  We find that major near-axis mergers heat cores most efficiently and at early epochs, such events (although between much lower mass systems than we study) are expected to be more common.  Recently \citet{Burnsetal07} have argued that early major mergers prevent compact cool cores from forming.  Systems having experienced a greater proportion of such events may have altered star formation histories as a result, yielding the observed differences in metallicity profiles between compact and extended core systems.  On the other hand, as noted by \citet{McCarthyetal07}, the virtual absence of extended core systems in all other studies involving cosmological simulations is compelling support for the idea that mergers alone are not responsible for the full range of observed cluster states.  Clearly, further investigation is required to enable us to select between competing possibilities.  Towards this goal, we should keep a keen eye-out for relaxed EWC systems with $t_{cool}>5$\Gyrs.  Such a system would be a strong candidate for a cluster that has been significantly heated by non-gravitational processes and would shed a great deal of light on these issues.

%-- SUMMARY --
\section{Summary and conclusions}\label{sec-summary}
We use the simulations presented in \citet{P06} to examine the effects of mergers on the properties of cool cores in X-ray clusters.  Our results suggest a natural scheme for classifying the morphology of clusters based on their central surface brightness and entropy profiles and we show that this scheme effectively captures the diversity of cluster characteristics revealed by recent high-resolution X-ray observations.  Our study illustrates how these morphologies naturally emerge in the context of hierarchical clustering.  For our present purposes we have not included the effects of pre-heating or post-heating by AGN.  We find:

\begin{itemize}
\item Systems exist in three dominant morphologies: those with compact cores and central temperatures which are cool (CCC systems) or warm (CWC systems) and extended warm core (EWC) systems.  Members of each population can appear either relaxed or disturbed.
\item Examining the temporal evolution of the morphology of our systems, we find that during merger events, CCC systems are only disturbed by direct collisions with incoming cores.  This can occur in head on collisions or during second pericentric passage in off-axis mergers.
\item Our merger remnants are generally heated to warm core states by the time they relax.  In cases where they relax to compact core structures, they typically recover cool cores (\ie\ return to CCC states) after $\sim 3$\Gyrs.  The central entropy of a merger remnant principally determines whether the system initially relaxes to a compact or extended core state.
\item A number of highly non-linear, strongly coupled processes are responsible for setting the central entropy of the remnant.  The outcome is very sensitive to the initial mass ratio and orbits of the interacting systems.  A heuristic overview of our simulation results are as follows:
\begin{itemize}
\item In on-axis cases, 3:1 and 10:1 mergers paradoxically result in a more heated remnant core than the 1:1 mergers.  This is due to enhanced heating from the penetration of the secondary core through the primary core and reduced cooling efficiencies in the low-density transient structures which result.  In 1:1 head-on cases the cores inelastically collide and remain in high density structures.  This reduces the efficiency of entropy production by shocks and increases entropy loss through radiative cooling.
\item Of the off-axis mergers we have studied, the case which generates the most significantly heated remnant core is a near-axis 3:1 merger.  In this case, the majority of core heating is due to the accretion of a low-entropy stream of material formed from the disruption of the secondary core.
\end{itemize}
\item Since $t_{dyn}\ll t_{cool}$ for all our relaxing merger remnant cores, compression prevents their core temperatures from falling until after they relax to the compact states allowed by their remnant central entropies.  This naturally explains the population of relaxed CWC systems observed in recent \Chandra\ and \XMM\ observations without any need to invoke AGN feedback.  We do not, however, exclude the possibility that extreme AGN outbursts may on rare occasions transform compact cold cores into compact warm cores.
\item The morphological segregation in scaling relations noted by \citet{MBBPH} is qualitatively reflected in the results of our mergers: EWC systems preferentially lie towards the low-luminosity/high-temperature side of the scatter in the $L_x-T_x$ plane.  However, as noted in Poole07, the typical cluster mergers we have studied (3:1 and 10:1 cases) do not produce systems with sufficiently high central entropies to account for the most under-luminous EWC systems.
\item In the $\beta-r_c$ plane, several trends in the observations can be naturally accounted for by mergers, but some can not.  In particular, our mergers nicely account for the observed dependence on $r_c$ of the high-value limit of $\beta$, with CCC systems lying at low values of $r_c$ and disturbed EWC systems lying at high values of $r_c$.  They also account for the observed tail of disturbed systems at the largest observed values of $\beta$.  They can not account for extended core systems with low-$\beta$ and long cooling times (\ie\ the majority of extended core systems).
\item Mergers do not efficiently mix the ICM.  Mergers between systems which initially host central metallicity gradients produce merger remnants with metallicity gradients only slightly reduced in central amplitude and slope than the initial systems.
\end{itemize}

Our results thus pose a significant challenge to a long held hypothesis in the study of X-ray clusters: that extended core systems are the remnants of mergers between relaxed compact core systems.  They suggest that compact cool core systems are remarkably stable to disturbances from cluster mergers.  Consequently, to explain the observed spectrum of X-ray cluster morphologies, further study of mechanisms capable of preventing the formation of compact cool cores is needed.

%-- APPENDIX --
%\begin{figure*}
%\begin{minipage}{175mm}
%\begin{center}
%\leavevmode \epsfysize=8cm \epsfbox{density_r_residual.ps} \epsfysize=8cm \epsfbox{temperature_r_residual.ps}
%\caption{Densities and temperatures of gas particles in an isolated $10^{14.5}M_{\odot}$ system after being evolved for 1 and 4 \Gyrs.  Blue crosses indicate particles for a run using our standard resolution (every other particle is plotted) and red points a $4\times$ mass resolution run (one-in-eight particles are plotted).  The heavy black dashed lines indicate power law fits to the $4\times$ mass resolution particles over the range indicated by the vertical dotted lines.  The bottom panels illustrate the residual after subtracting this power law, with short dashed lines indicating $\pm10\%$ residuals.}
%\label{fig-residuals}
%\end{center}
%\end{minipage}
%\end{figure*}
%
%\input{appendix}

%-- THANKS --
\section*{Acknowledgements}
We would like to thank Ken Cavagnolo and Megan Donahue for providing information about their unpublished catalogue of Chandra observations.  We would also like to thank Michael Balogh, Neal Katz and Alastair Edge for stimulating discussions and insightful comments.  GBP acknowledges the financial support of the Australian Research Council.  IGM acknowledges support from a NSERC Postdoctoral Fellowship and a PPARC rolling grant for extragalactic astronomy and cosmology at the University of Durham.  AB acknowledges support from NSERC through the Discovery Grant program.  MF acknowledges support from NSERC, NASA, and NSF.

%-- REFERENCES --

\label{lastpage}

\end{document}